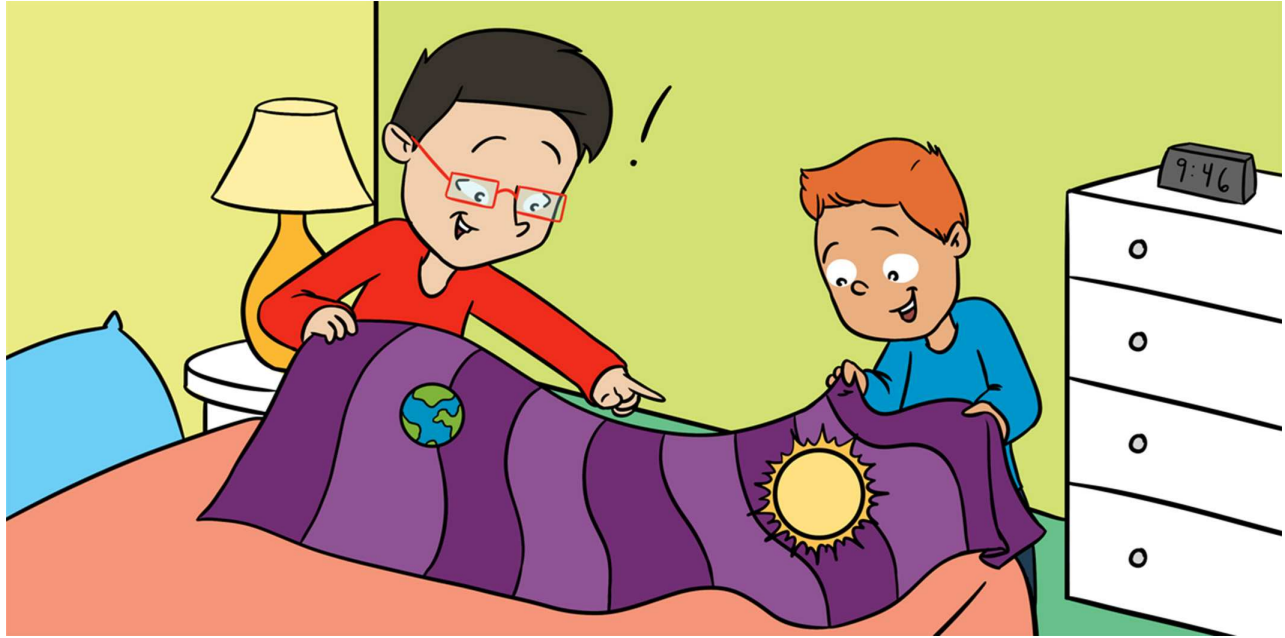

# CATCHING GRAVITATIONAL WAVES WITH A GALAXY-SIZED NET OF PULSARS

*Stephen R. Taylor* \*

*TAPIR Group (Theoretical AstroPhysics Including Relativity & Cosmology), California Institute of Technology, Pasadena, CA, United States*

**YOUNG REVIEWERS:**

**DUARTE HIGH SCHOOL**
AGES: 15–18

**LA CANADA HIGH SCHOOL**
AGES: 16–18

Until recently, the only way to observe the Universe was from light received by telescopes. But we are now able to measure gravitational waves, which are ripples in the fabric of the Universe predicted by Albert Einstein. If two very dense objects (like black holes) orbit each other closely, they warp space and send out gravitational waves. For black holes that are similar in mass to the Sun, scientists use the LIGO detector on Earth. But for the biggest black holes in the Universe (billions of times more massive than the Sun), scientists monitor a net of rapidly-spinning neutron stars (called pulsars) across the Milky Way. Any gravitational wave passing by will change how long radio signals from these pulsars take to get to Earth. The NANOGrav Collaboration monitored 34 of these pulsars over 11 years, in an attempt to detect gravitational waves from giant black holes.

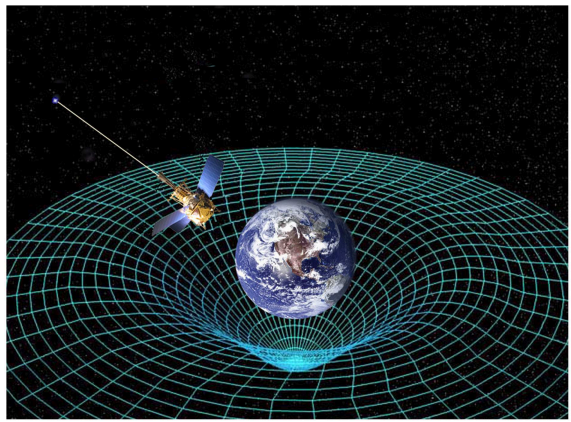

**Figure 1**

**Figure 1**
The Earth makes a dip in the fabric of space, as do all other objects. The more massive the object, the bigger the dip. This is what causes the pull of gravity, allowing other objects (like satellites) to orbit. Image credit: NASA, https://www.nasa.gov/mission_pages/gpb/gpb_012.html.

## INTRODUCTION

You may have heard the phrase, "What goes up, must come down." But why is this? It is because everything (you, me, rockets, raindrops) is being pulled toward the Earth by gravity. The strength of this pull is stronger for bigger objects but gets weaker for objects that are far apart. Rockets are only able to escape the pull of the Earth's gravity by burning huge amounts of fuel to create an upwards "push" force that is stronger than gravity's "pull."

What is amazing is that gravity is not special to Earth; all objects in the Universe feel the same type of gravity force, which holds stars, solar systems, and galaxies together. Even 100 years ago, astronomers knew how to predict where objects in the solar system should appear in the night sky. But they did not know what gravity really was. This is when Albert Einstein had a very big idea: what we call gravity is not really a "pull force" like being attached to a rope. In Einstein's theory, space is more like a stretchy blanket than a hard table; every object makes a dip in this blanket, but bigger objects make bigger dips. So, a star like the Sun makes a big dip, and the Earth makes a much smaller dip (see Figure 1). The Earth is moving slowly enough that it is falling in toward the Sun's dip, but also fast enough that it does not spiral in immediately: this is called an orbit.

## GRAVITATIONAL WAVES

Since space is more like a blanket than a hard table, it can squeeze and stretch to make waves. You can think of standing in a circle with a group of friends, each holding a part of a round blanket. If your friend on the opposite side of the blanket shakes their part up and down, then your part will shake, too. Even though your part of the blanket seemed to shake at the same time as your friend's, it took a tiny amount of time for the wave to reach you. The same thing happens in space, except

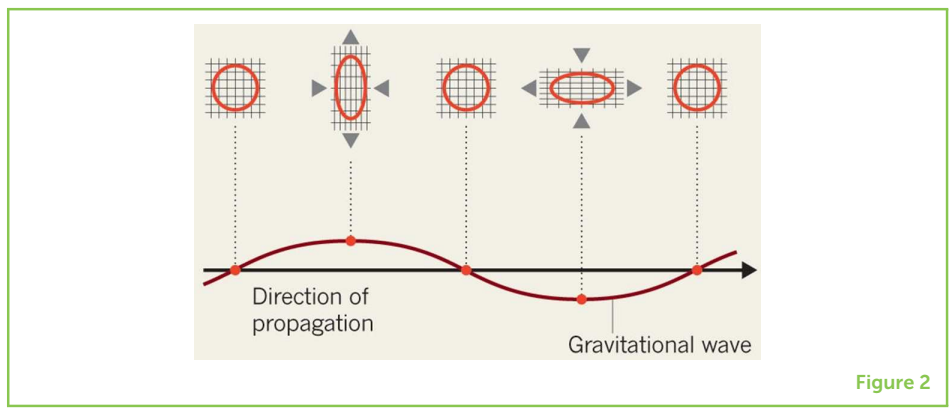

Figure 2

# Figure 2

Gravitational waves travel at the speed of light. They stretch and squeeze the fabric of space in the plane perpendicular to the direction of travel. Image credit: Image credit: Markus Pössel/Einstein-online.info.

# GRAVITATIONAL WAVE

A ripple in space and time caused by accelerating massive objects.

# LIGO

Stands for Laser Interferometer Gravitational-wave Observatory. This is a pair of large gravitational-wave detectors located in Hanford, Washington and Livingston, Louisiana. They detected gravitational waves for the first time in 2015.

# VIRGO

A similar gravitational-wave detector to LIGO, but located in Italy.

# NEUTRON STAR

A very dense object formed when a normal star runs out of fuel and collapses in on itself due to gravity. The collapse is halted by the pressure of neutrons resisting against gravity.

# BLACK HOLE

When a star cannot prevent total collapse caused by gravity, a black hole is formed. It is a point of infinite density. Not even light can escape from it.

that space is really quite stiff, so it takes gigantic objects to make a big wave. If there are two objects that are moving around each other in an orbit, they can shake space enough that they make waves. These waves are called **gravitational waves**, and they move at the speed of light.

As a gravitational wave travels, it stretches and squeezes space. But this stretching and squeezing does not happen along the direction that the wave is moving. Instead, the stretching and squeezing happens perpendicular to the direction that the gravitational wave is moving (see Figure 2). Scientists have recently used specialized equipment to build big experiments that are sensitive enough to measure this stretching and squeezing, such as **LIGO** [1] and **Virgo** [2]. But, as mentioned above, space is very stiff, so we can still only measure waves from special types of dense astronomical objects that are very close together. These objects are called **neutron stars** and **black holes**.

## NEUTRON STARS AND BLACK HOLES

Like a rocket, stars have to burn huge amounts of fuel to create an outwards push that stops gravity from collapsing the star. When a star runs out of fuel, it begins to collapse inwards, but can save itself from total collapse twice along the way. The first time is when the atoms inside the star cannot be squeezed any closer together. This saves the star for a little while, but if the star is big enough, then gravity is so strong that it squeezes all the atoms together. The last chance for this star to survive collapsing due to gravity is when the tiny building blocks of the atom, known as neutrons, stop themselves from being squeezed together. This creates what is called a neutron star. The matter in a neutron star is packed so tightly together that a spoonful would weigh as much as a skyscraper! Finally, if we have a really huge star, then not even the neutrons resistance to being squeezed can stop total collapse. The total collapse of a star makes a black hole. Black holes are not made of matter; they are made of gravity itself, creating a dip

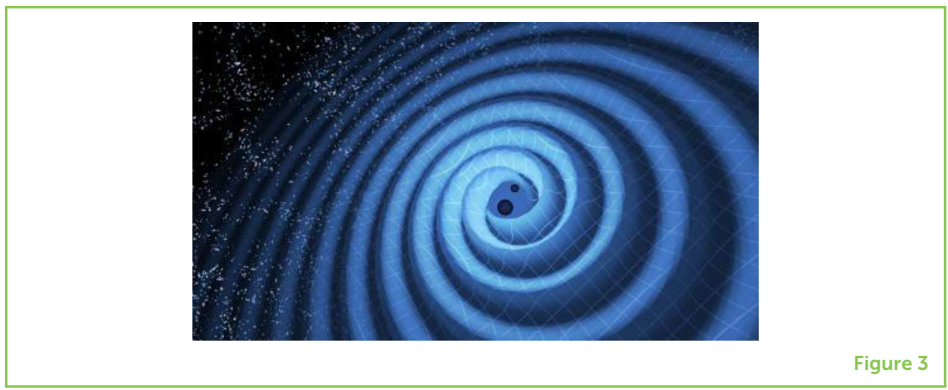

Figure 3

### Figure 3

Two black holes that orbit each other will make ripples in space, called gravitational waves. These waves carry energy away from the orbit of the black holes, causing the two to eventually merge. Image credit: LIGO/T. Pyle, https://www.ligo.caltech.edu/image/ligo20160615f.

in the fabric of space that is so deep that not even light can escape its pull; this is why they are called "black" holes.

## MEASURING THE MOST MASSIVE BLACK HOLES IN THE UNIVERSE

Sometimes black holes can pair up together and orbit one another (see Figure 3). When the two black holes get very close in their orbit, they can emit gravitational waves. A team of scientists called **NANOGrav** (short for the North American Nanohertz Observatory for Gravitational waves) [3] has been hunting for gravitational waves emitted by pairs of the most massive black holes in the Universe. These black holes can be as massive as a billion Suns and are only found at the centers of giant galaxies. Over the history of the Universe, galaxies have collided together, making even bigger galaxies. In these collisions, the black holes from the galaxy centers paired up, sending out gravitational waves that have a period (the time between each wave peak) of years to decades. Since black holes do not emit any light, the only way to detect them is with gravitational waves. Measuring gravitational waves is a radically new way of observing the Universe and these measurements will tell us more about the true nature of gravity.

### NANOGRAV

Stands for North American Nanohertz Observatory for Gravitational waves. This is a group of scientists in the USA and Canada who hunt for gravitational waves using PTAs.

## PULSAR TIMING ARRAYS

NANOGrav searches for gravitational waves using something called a pulsar-timing array **(PTA)** [4]. A pulsar is a special kind of neutron star that spins around very quickly (as quickly as hundreds of times each second) and shoots out beams of radio waves (see the left side of Figure 4). Whenever a pulsar spins around, it shoots the radio wave beam toward Earth, which we measure as a radio "pulse." This is like a lighthouse, which is always shining its light at night time, but we only see the light when it swings around to face us. Pulsars are very reliable; we can very accurately predict when the radio pulses will arrive at Earth. This means we can use pulsars like a stopwatch, with which we mark the passage of time by the number of radio pulses that have been

### PTA

Stands for Pulsar Timing Array. Pulsars are a special type of spinning neutron star that send out radio beams. When gravitational waves come into our galaxy, they cause changes in the arrival time of the radio beams from pulsars. We look at many pulsars to confirm that these changes are caused by gravitational waves and not noise.

# Figure 4

Pulsars (left) are very special types of neutron stars that spin around as fast as hundreds of times each second. They regularly send out radio beams that can be measured on Earth very precisely. By observing many pulsars across the Milky Way galaxy (right), we can look for any common changes to the arrival time of radio pulses. These changes may be caused by gravitational waves passing through the Milky Way. The figure on the right shows a cartoon illustration of four pulsars in the Milky Way (gray regions with yellow stars) beaming radio waves toward our Solar system. Image credit: Tonia Klein/Jeffrey Hazboun/The NANOGrav Physics Frontier Center.

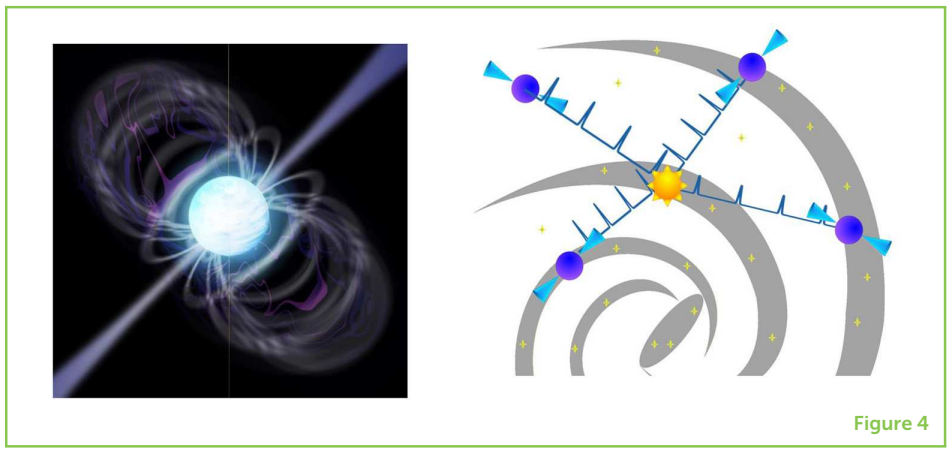

Figure 4

observed from a pulsar. Pulsars are great stopwatches that stay reliable over many years.

If a gravitational wave crosses the space between Earth and a pulsar, it will stretch and squeeze that space. If space is stretched, it will take longer than expected for the radio beam to reach us; the pulse will arrive late! The opposite is true if space is squeezed, because the radio pulse will arrive earlier than expected. We can subtract our predictions of when the radio pulses *should* arrive from our real observations, and look at the difference. That difference could be due to gravitational waves! NANOGrav (and other teams in Europe, Australia, India, South Africa, and China) have been using radio telescopes to observe lots of pulsars to make a big "net" that can catch gravitational waves.

## NANOGRAV'S NEW DISCOVERY

In NANOGrav's most recent hunt [5], we made a net out of 34 of these pulsars that have been watched by astronomers every couple of weeks over the last 11 years. We did not find any gravitational waves, but we also know that our signals take a long time to stick out above all of the noise and weird stuff that can affect pulsars. Even though we have not detected anything yet, we think it will only be another 3 years, or possibly 7 years at the most, before we do [6]. We may not have seen any waves yet, but this absence of waves has allowed us to disprove predictions made by other scientists who thought we should have seen something by now. Our data will help those scientists to revise and update their predictions. The results we have obtained also help us understand how often massive black holes merge together in the Universe. We also found that gravitational waves with periods of 1 year cause stretches and squeezes to space that are very, very tiny—so small that the change they cause to the size of the Earth is only about 10 times the width of a human DNA strand (see Figure 5)!

#### Figure 5

NANOGrav has put limits on the size of the fractional squeezing caused by gravitational waves. Everything in the red region has been ruled out by our analysis of 11 years of pulsar-timing data. The green region shows the range of gravitational-wave predictions made by other scientists.

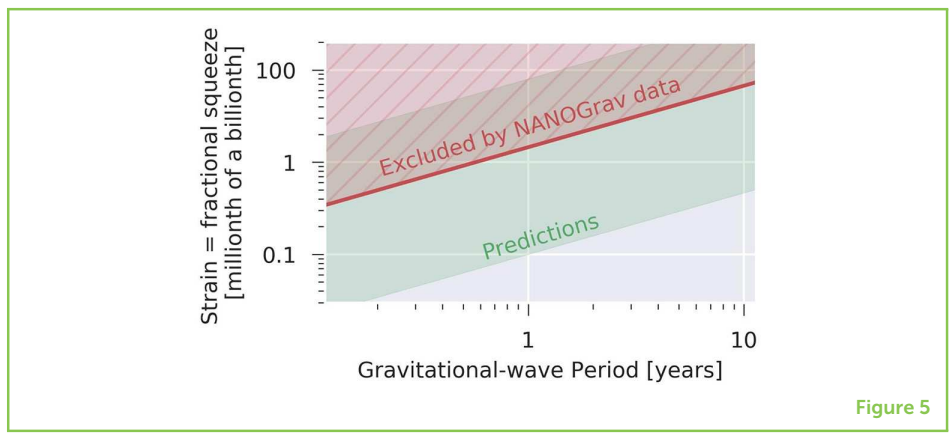

Figure 5

## THE FUTURE

We expect the future of gravitational-wave astronomy to be very exciting, allowing us to peer into parts of the Universe that other telescopes cannot see. The detection of gravitational waves by pulsar-timing arrays in the near future will be a huge discovery. In NANOGrav, we are checking for other types of gravitational waves in this net of pulsars and will report on them over the next year. We are also constantly searching for new pulsars to fill holes in our net, so that we are better able to find gravitational waves. Our results so far have been fascinating, and we are preparing for the day quite soon when we can tell the world that we have seen gravitational waves from the most massive black holes in the entire Universe!

## ORIGINAL SOURCE ARTICLE

Arzoumanian, Z., Baker, P. T., Brazier, A., Burke-Spolaor, S., Chamberlin, S. J., Chatterjee, S., et al., The NANOGrav Collaboration. 2018. The NANOGrav 11 year data set: pulsar-timing constraints on the stochastic gravitational-wave background. *Astrophys. J.* 859:47. doi: 10.3847/1538-4357/aabd3b

## FUNDING

ST is funded by the NANOGrav Physics Frontier Center, which is supported by NSF award number 1430284. The Green Bank Observatory is a facility of the National Science Foundation operated under cooperative agreement by Associated Universities, Inc. The Arecibo Observatory is a facility of the National Science Foundation operated under cooperative agreement by the University of Central Florida in alliance with Yang Enterprises, Inc. and Universidad Metropolitana.

**SUBMITTED:** 07 December 2018; **ACCEPTED:** 23 May 2019;
**PUBLISHED ONLINE:** 07 June 2019.

**EDITED BY:** Joey Shapiro Key, University of Washington Bothell, United States

**CITATION:** Taylor SR (2019) Catching Gravitational Waves With a Galaxy-Sized Net of Pulsars. Front. Young Minds 7:80. doi: 10.3389/frym.2019.00080

**CONFLICT OF INTEREST STATEMENT:** The author declares that the research was conducted in the absence of any commercial or financial relationships that could be construed as a potential conflict of interest.

## YOUNG REVIEWERS

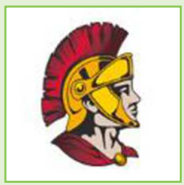

### DUARTE HIGH SCHOOL, AGES: 15–18
We are Physics students in Mr. Traeger's Period 2 and 3 Physics classes at Duarte High School in Duarte, CA. We are interested in furthering science through the peer review process.

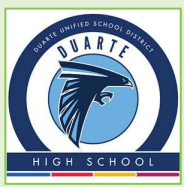

### LA CANADA HIGH SCHOOL, AGES: 16–18
We are AP Physics students in Ms. Walsh's Period 5 Physics class at La Canada High School in La Canada, CA. We are interested in learning more about the scientific process through the peer review.

## AUTHOR

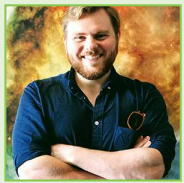

**STEPHEN R. TAYLOR**
I am currently a post-doctoral research fellow at the California Institute of Technology, a visiting scientist at NASA's Jet Propulsion Laboratory, and from Fall 2019 will be an Assistant Professor at Vanderbilt University. I am an astrophysicist, studying what we can learn about the Universe by detecting gravitational waves. These waves come from colliding black holes and neutron stars throughout the Universe, and give information that is different to what we measure with telescopes. In my free time, I like to read, hike, watch movies, and spend time with family. *srtaylor@caltech.edu